\documentclass[aps,prl,twocolumn,showpacs,preprintnumbers,amsmath,amssymb]{revtex4-1}
\usepackage[dvipdfmx]{graphicx}

 \def\frac#1#2{{#1\over #2}}

 \def\s{\sqrt}

\def\be{\begin{equation}}
\def\ee{\end{equation}}
\def\ba{\begin{eqnarray}}
\def\ea{\end{eqnarray}}

 \def\f {\frac}
 \def\ti{\tilde}
 \def\ap{\alpha}

 \def\ddd{\cdot\cdot\cdot}
 \def\no{\nonumber \\}

 \def\la{\langle}
 \def\lb{\rangle}

 \def\vp{\varphi}

\usepackage{color}
\usepackage{graphicx}
\usepackage{dcolumn}
\usepackage{bm}

\begin{document}

\title{Entanglement Wedges from Information Metric in
Conformal Field Theories}
YITP-19-80 ; IPMU19-0114
\author{Yuki Suzuki$^{a}$,  Tadashi Takayanagi$^{b,c}$ and Koji Umemoto$^{b}$}

\affiliation{$^a$ Faculty of Science, Kyoto University,\\
Kitashirakawa Oiwakecho, Sakyo-ku, Kyoto 606-8502, Japan}

\affiliation{$^b$Center for Gravitational Physics,\\
Yukawa Institute for Theoretical Physics,
Kyoto University, \\
Kitashirakawa Oiwakecho, Sakyo-ku, Kyoto 606-8502, Japan}

\affiliation{$^{c}$Kavli Institute for the Physics and Mathematics
 of the Universe (WPI),\\
University of Tokyo, Kashiwa, Chiba 277-8582, Japan}

\date{\today}

\begin{abstract}
We present a new method of deriving the geometry of entanglement wedges 
in holography directly from conformal field theories (CFTs).
We analyze an information metric called the Bures metric of reduced density matrices for locally excited states. This measures distinguishability of states with different points excited.
For a subsystem given by an interval, we precisely reproduce the expected entanglement wedge for two dimensional holographic CFTs
from the Bures metric, which turns out to be proportional
to the AdS metric on a time slice. On the other hand,  for free scalar CFTs, we do not find any sharp structures like entanglement wedges. When a subsystem consists of disconnected two intervals we manage to reproduce the expected entanglement wedge from holographic CFTs with correct phase transitions, up to a very small error,
from a quantity alternative to the Bures metric.
\end{abstract}

\maketitle

{\bf 1. Introduction}

An important and  fundamental question in the anti-de Sitter space/conformal field theory
 (AdS/CFT) correspondence \cite{Ma} is
``Which region in AdS corresponds to a given subregion $A$ in a CFT ?''.
The answer to this question has been argued to be the entanglement wedge $M_A$ \cite{EW1,EW2,EW3},
i.e. the region surrounded by the subsystem $A$ and the extremal surface $\Gamma_A$ whose area gives the
holographic entanglement entropy \cite{RT,HRT}.  Here the reduced density matrix $\rho_A$ on the subregion
$A$ in a CFT gets dual to the reduced density matrix $\rho^{bulk}_{M_A}$ on the entanglement wedge
$M_A$ in the dual AdS.

Normally this bulk-boundary subregion duality is explained
by combining several ideas: the gravity dual of bulk field operator (called HKLL map \cite{HKLL}),
the quantum corrections to holographic entanglement entropy \cite{FLM,JLMS} and
the conjectured connection between AdS/CFT and quantum error correcting codes \cite{ADH,Dong:2016eik}.
However, since this explanation highly employs the dual AdS geometry and its dynamics from the beginning, 
it is not clear
how the entanglement wedge geometry emerges from a CFT itself. The main aim of this article is to derive the geometry of
entanglement wedge purely from CFT computations. We will focus on two dimensional (2d) CFTs for technical reasons.
The AdS/CFT argues that a special class of CFTs, called holographic CFTs, can have
classical gravity duals which are well approximated by general relativity. A holographic CFT is characterized by
a large central charge $c$ and very strong interactions, which lead to a large spectrum gap \cite{He,Hartman:2014oaa}. Therefore
we expect that the entanglement wedge geometry is available only when we consider holographic CFTs.
Our new framework will explain how entanglement
wedges emerge from holographic CFTs.

For this purpose we consider a locally excited state in a 2d CFT, created by acting a primary operator $O(w,\bar{w})$ on the vacuum.
We focus on the 2d CFT which lives on an Euclidean complex plane R$^2$, whose coordinate is denoted by
$(w,\bar{w})$ or equally $(x,\tau)$ such that $w=x+i\tau$.
By choosing a subsystem $A$ on the $x$-axis,  we define the reduced density matrix on $A$, tracing out its complement $B$:
\ba
\rho_A(w,\bar{w})=\mbox{Tr}_B\left[O(w,\bar{w})|0\lb \la 0|O^\dagger(\bar{w},w)\right], \label{redb}
\ea
first introduced in \cite{NNT} to study its entanglement entropy.

\begin{figure}
  \centering
  \includegraphics[width=4cm]{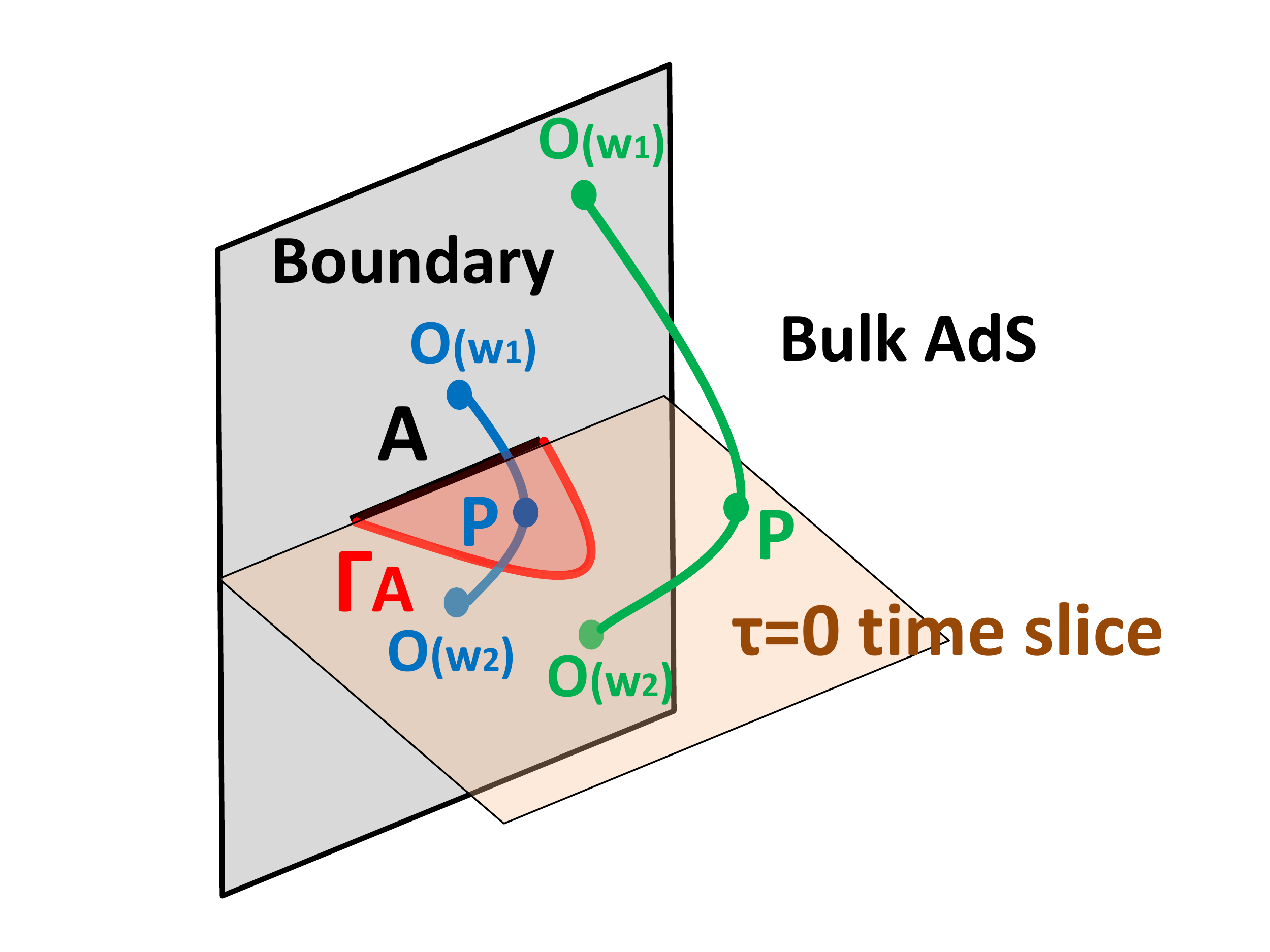}
  \caption{A sketch of entanglement wedge $M_A$ for an interval $A$ in AdS$_3/$CFT$_2$ and holographic computations of two point functions dual to geodesics. The blue (or green) geodesic does (or does not) intersect with $M_A$.}
\label{fig:EW}
  \end{figure}

We assume that the (chiral and anti chiral) conformal dimension $h$ of $O$ satisfies $1\ll h\ll c$.
In this case, we can neglect its back reaction in the gravity dual and can approximate the two point function
$\la O(w_1,\bar{w}_1) O^\dagger(w_2,\bar{w}_2)\lb$ by the geodesic length in the gravity dual
between the two points  $(w_1,\bar{w}_1)$ and $(w_2,\bar{w}_2)$ on the boundary $\eta\to 0$ of the Poincare AdS$_3$
\be
ds^2=\eta^{-2}(d\eta^2+dwd\bar{w})=\eta^{-2} (d\eta^2+dx^2+d\tau^2), \label{po}
\ee
where we set the AdS radius one.
Therefore, by projecting on the bulk time slice $\tau=0$, the state $\rho_A(w,\bar{w})$ is expected to be dual to a bulk excitation at a bulk point $P$ defined by the intersection between the time slice $\tau=0$ and the geodesic, as depicted in Fig.\ref{fig:EW}.

Now we are interested in how we can distinguish the two states: $\rho_A(w,\bar{w})$ and  $\rho_A(w',\bar{w}')$ when $w\neq w'$,
created by the same operators.
They are dual to bulk states with two different points excited. On the time slice $\tau=0$, the location of bulk excitations are given by
$(\eta,x)=(\tau,x)$ and $(\eta,x)=(\tau',x')$.
The entanglement wedge reconstruction argues we cannot distinguish the two excited bulk states when both excitations are
outside of $M_A$, while we can distinguish them if at least one of them is inside of $M_A$.

A useful measure of distinguishability between two density matrices $\rho$ and $\rho'$ is
the Bures distance $D_B$, defined by (refer to  e.g. \cite{Hayashi})
\ba
&& D_B(\rho,\rho')^2=2(1-\mbox{Tr}[\s{\s{\rho}\rho'\s{\rho}}]), \label{disb}
\ea
Moreover we can define the information metric when the density matrix is parameterized by continuous
valuables $\lambda^i$, denoted by $\rho(\lambda)$:
\ba
D_B(\rho(\lambda+d\lambda),\rho(\lambda))^2\simeq G_{ij}d\lambda^i d\lambda^j\equiv dD_B^2,
\ea
where $d\lambda_i$ are infinitesimally small. This metric $G_{ij}$ is called the Bures metric,
which measures the distinguishablility between nearby states.

The quantum version of Cramer-Rao theorem \cite{Hel} tells us that when we try to estimate the value of
$\lambda_i$ from physical measurements, the errors of the estimated value is bounded by the inverse of the Bures metric as follows
\ba
\la\la \delta \lambda^i\delta\lambda^j \lb\lb \geq (G^{-1})^{ij}.
\ea
This shows as the Bures metric gets larger, the errors due to quantum fluctuations
 get smaller.

As an exercise, consider the case where $A$ covers  the total system, where
$\rho_A(w,\bar{w})$ becomes a pure state $|\phi(w)\lb\la\phi(w)|$.
The Bures distance $D_B$ is simplified as
\ba
&& D_B(|\phi\lb,|\phi'\lb)^2=2(1-|\la\phi(w)|\phi(w')\lb|), \no
&& |\la\phi(w)|\phi(w')\lb|=|w-\bar{w}|^{2h}|w'-\bar{w}'|^{2h}|w-\bar{w}'|^{-4h}.
\ea
This leads to the Bures metric
\be
dD_B^2= \frac{h}{\tau^2}(d\tau^2+dx^2).  \label{hypp}
\ee
In this way, the information metric is proportional to the actual metric in the gravity dual (\ref{po})
on the time slice $\tau=0$. This coincidence is very natural because the
distinguishability should increase as the bulk points are geometrically separated and
was already noted essentially in \cite{MNSTW}. However, this result is universal for any 2d CFTs.
Soon later we will see this property largely changes for reduced density matrices, where
results crucially depend on CFTs. We will be able to 
find the entanglement wedge structure only for holographic CFTs.

Before we go on, let us mention that for technical conveniences, we often calculate (introduced in  \cite{Cardy:2014rqa})
\ba
I(\rho,\rho')=\frac{\mbox{Tr}[\rho\rho']}{\s{\mbox{Tr}[\rho^2]\mbox{Tr}[\rho'^2]}}, \label{cq}
\label{dca}
\ea
instead of $\mbox{Tr}[\s{\s{\rho}\rho'\s{\rho}}]$ to estimate distinguishability.
If $\rho=\rho'$ we find $I(\rho,\rho')=1$, while we have $0<I(\rho,\rho')<1$ 
when $\rho\neq \rho'$.\\

{\bf 2. Single Interval Case}

We choose the subsystem $A$ to be an interval $0\leq x\leq L$ at $\tau=0$.
The surface $\Gamma_A$ in the bulk AdS is given by the semi circle $(x-L/2)^2+\eta^2=L^2/4$.
Therefore if the entanglement reconstruction is correct,
the information metric should vanish if the intersection $P$ is outside of the entanglement wedge given by
\ba
|w-L/2|=L/2.  \label{outent}
\ea
In this example, the entanglement wedge is equivalent to the causal wedge \cite{HR}.

\begin{figure}
  \centering
  \includegraphics[width=5cm]{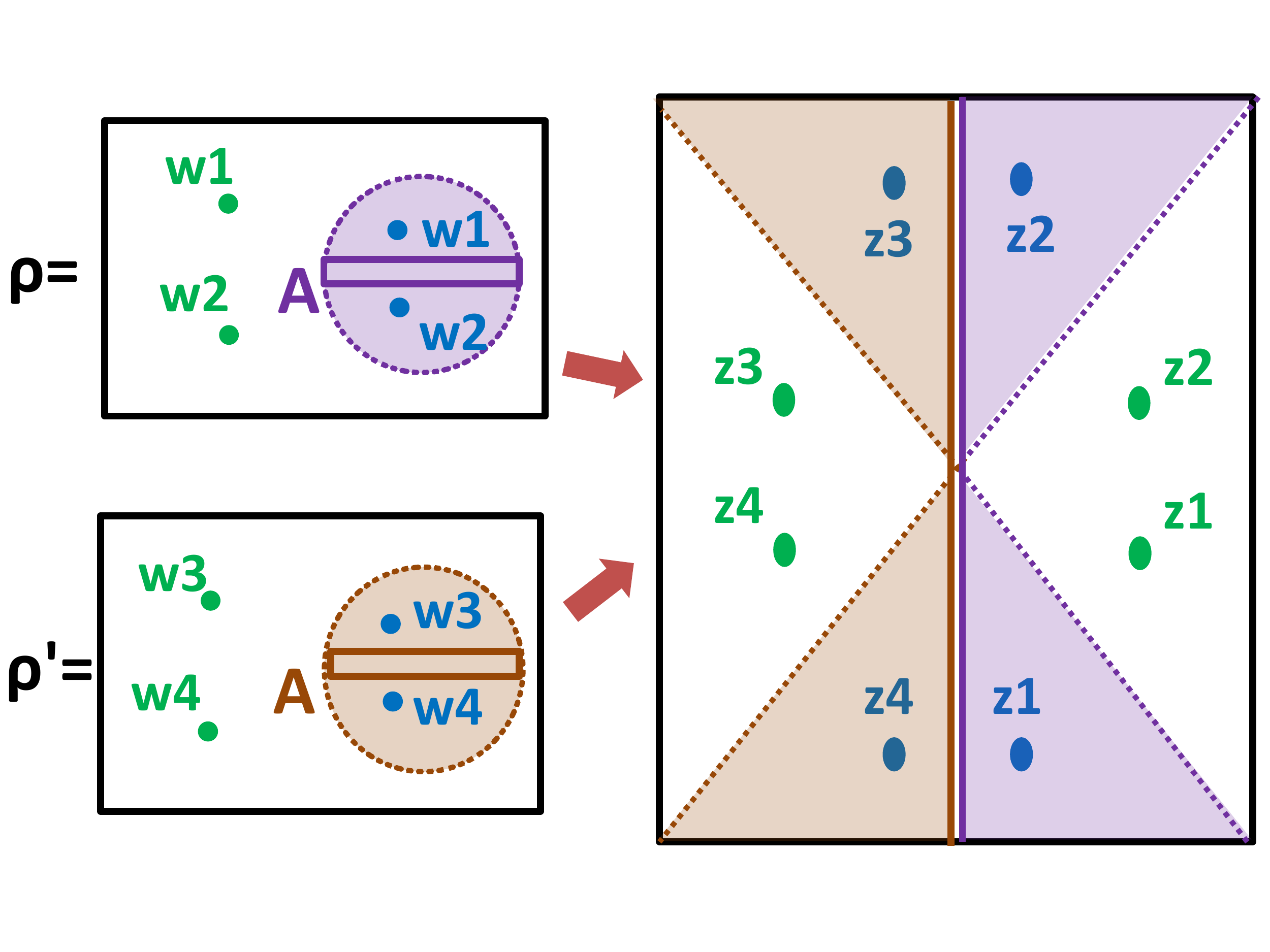}
  \caption{A sketch of  conformal transformation for the calculation of Tr$[\rho\rho']$. Green Points
(or bule points) are outside (or inside) of  the wedge (\ref{outent}).}
\label{fig:singlePO}
  \end{figure}

Let us start with the calculation of the quantity $I(\rho,\rho')$ defined by (\ref{dca}),
for $\rho=\rho_A(w,\bar{w})$ and  $\rho'=\rho_A(w',\bar{w}')$.
Since this calculation is essentially that of Tr$[\rho\rho']$,
we  perform the conformal transformation:
\ba
z^2=\frac{w}{w-L},  \label{sintr}
\ea
which maps two flat space path-integrals which produce $\rho(w,\bar{w})$ and $\rho(w',\bar{w}')$
into a single plane ($z-$plane). Refer to \cite{Nozaki,HNTW} for similar calculations
in the context of entanglement entropy of such states.
The insertion
points of the four primary operators on the $z-$plane are given by (remember $w=x+i\tau$)
\ba
&& z_1\!=\!\s{\frac{-x-i\tau}{L-x-i\tau}}(\equiv\! z), \ \  \ z_2\!=\!\s{\frac{-x+i\tau}{L-x+i\tau}}(\equiv\! \bar{z}),\no
&& z'_3\!=\!-\s{\frac{-x'-i\tau'}{L-x'-i\tau'}}(\equiv\! -z'), \  z'_4\!=\!-\s{\frac{-x'+i\tau'}{L-x'+i\tau'}}(\equiv\! -\bar{z}').
\nonumber
\ea
Refer to Fig.\ref{fig:singlePO} for this conformal mapping.
It is important to note that the boundaries of the wedges  (\ref{outent}) on the $w-$planes
 are mapped into the diagonal lines $z=\pm i \bar{z}$.

The quantity Tr$[\rho\rho']$ is expressed as a correlation function on the $z-$plane:
\ba
&& \mbox{Tr}[\rho\rho']\!=\!\left[\left|\frac{dz_1}{dw_1}\right|\left|\frac{dz_2}{dw_2}\right|
\left|\frac{dz'_3}{dw'_3}\right|\left|\frac{dz'_4}{dw'_4}\right|\right]^{2h}\!\!\cdot\!
\frac{H(z_1,z_2,z'_3,z'_4)\!\cdot \!Z^{(2)}}{(Z^{(1)})^2}, \no
&&
H(z_1\!,z_2,\!z'_3,\!z'_4)\!\equiv\!\frac{\la O^\dagger(z_1,\!\bar{z}_1)O(z_2,\!\bar{z}_2)
O^\dagger(z'_3,\!\bar{z}'_3)O(z'_4,\!\bar{z}'_4)\lb}{\la O^\dagger\!(w_1,\!\bar{w}_1)O\!(w_2,\!\bar{w}_2)\lb
\la O^\dagger\!(w'_3,\!\bar{w}'_3)O\!(w'_4,\!\bar{w}'_4)\lb},\nonumber
\ea
where $\la \ddd \lb$ denotes the normalized correlation function such that $\la 1\lb=1$ and
we also write the vacuum partition function on a $n$-sheeted complex plane by $Z^{(n)}$.
Finally we obtain
\ba
&& I(\rho,\rho')=\frac{F(z_1,z_2,z'_3,z'_4)}{\s{F(z_1,z_2,z_3,z_4)F(z'_1,z'_2,z'_3,z'_4)}},  \label{formi} \\
&& F(z_1\!,z_2,\!z'_3,\!z'_4)\!\equiv\! \la O^\dagger(z_1,\!\bar{z}_1)O(z_2,\!\bar{z}_2)
O^\dagger(z'_3,\!\bar{z}'_3)O(z'_4,\!\bar{z}'_4)\lb. \nonumber
\ea

\begin{figure}
  \centering
  \includegraphics[width=3cm]{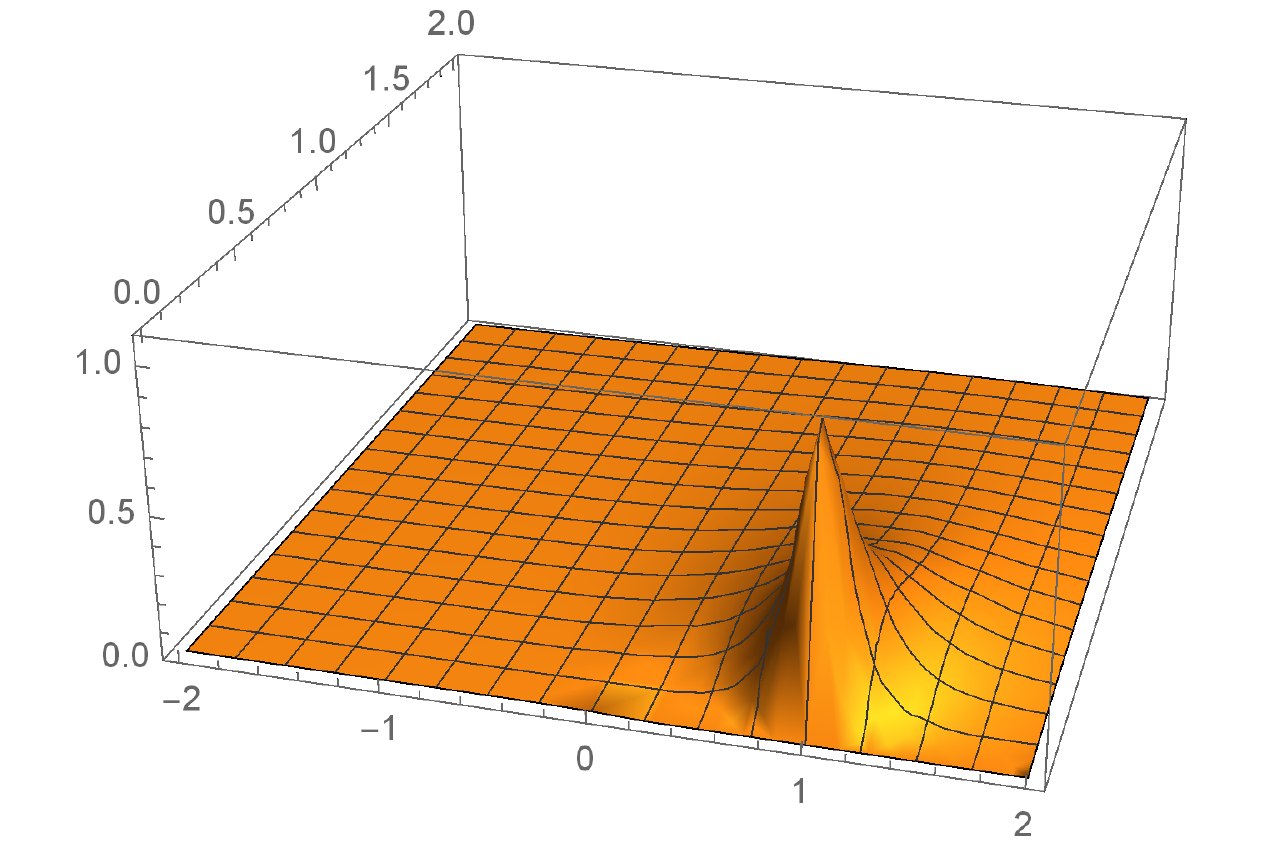}
 \includegraphics[width=3cm]{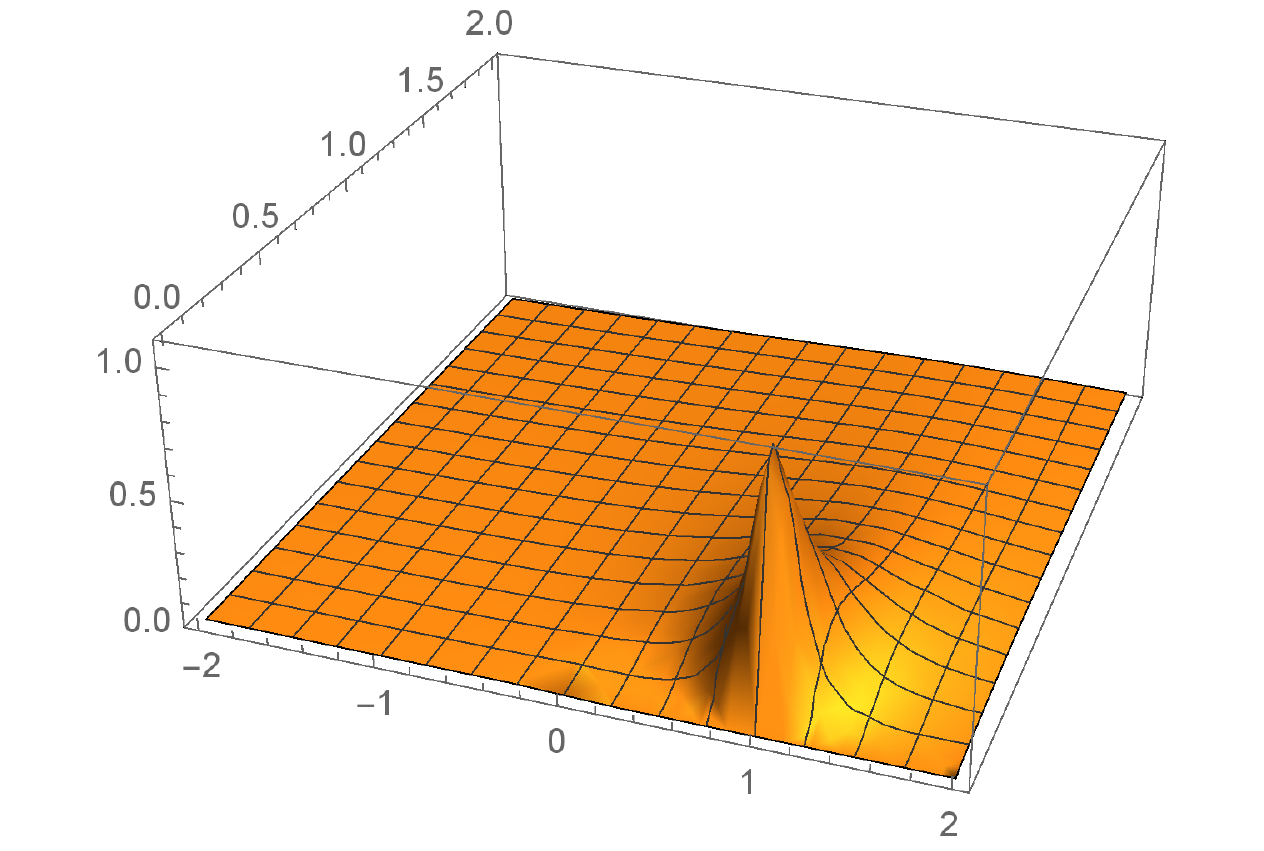}
\includegraphics[width=2.5cm]{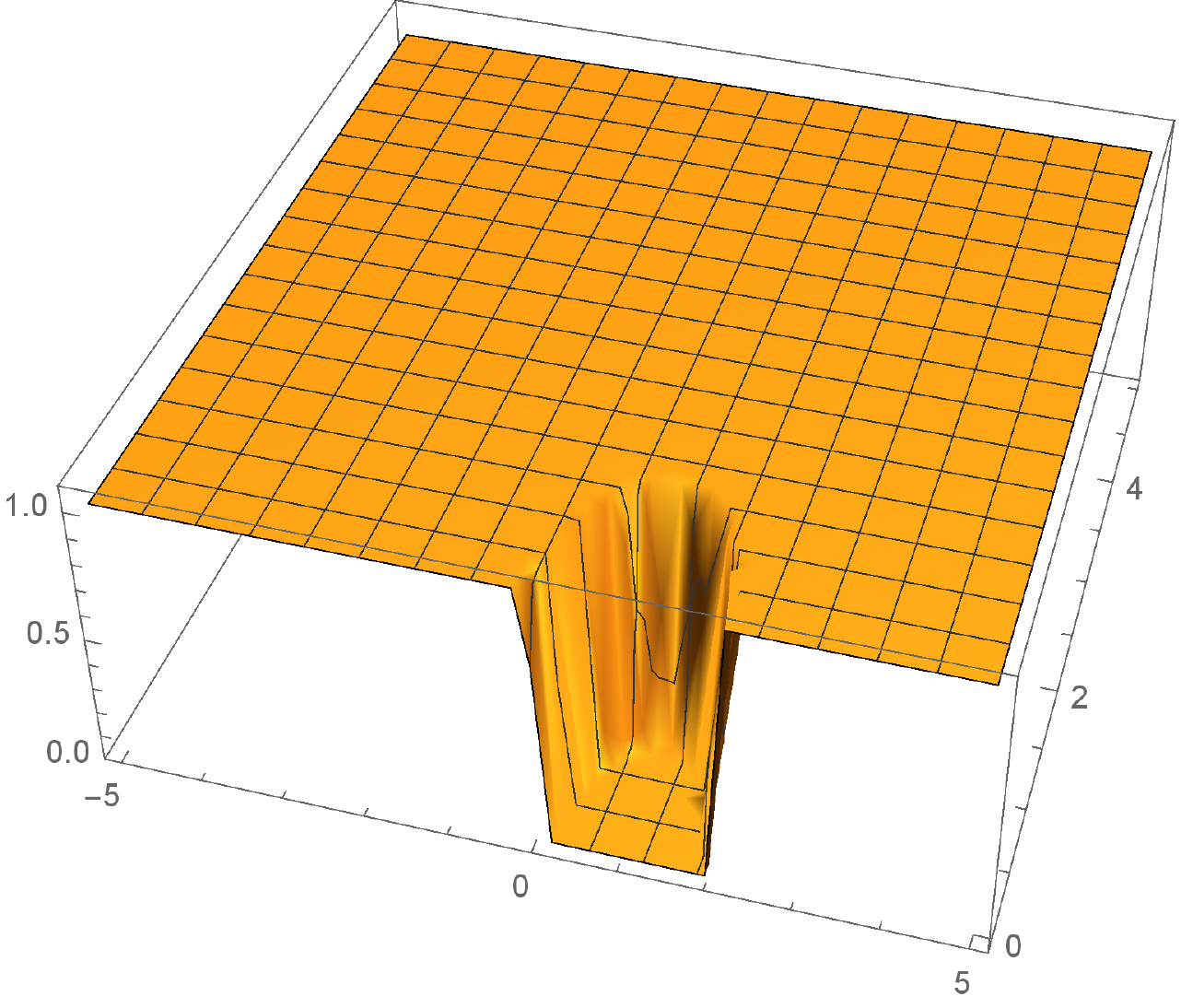}
 \includegraphics[width=2.5cm]{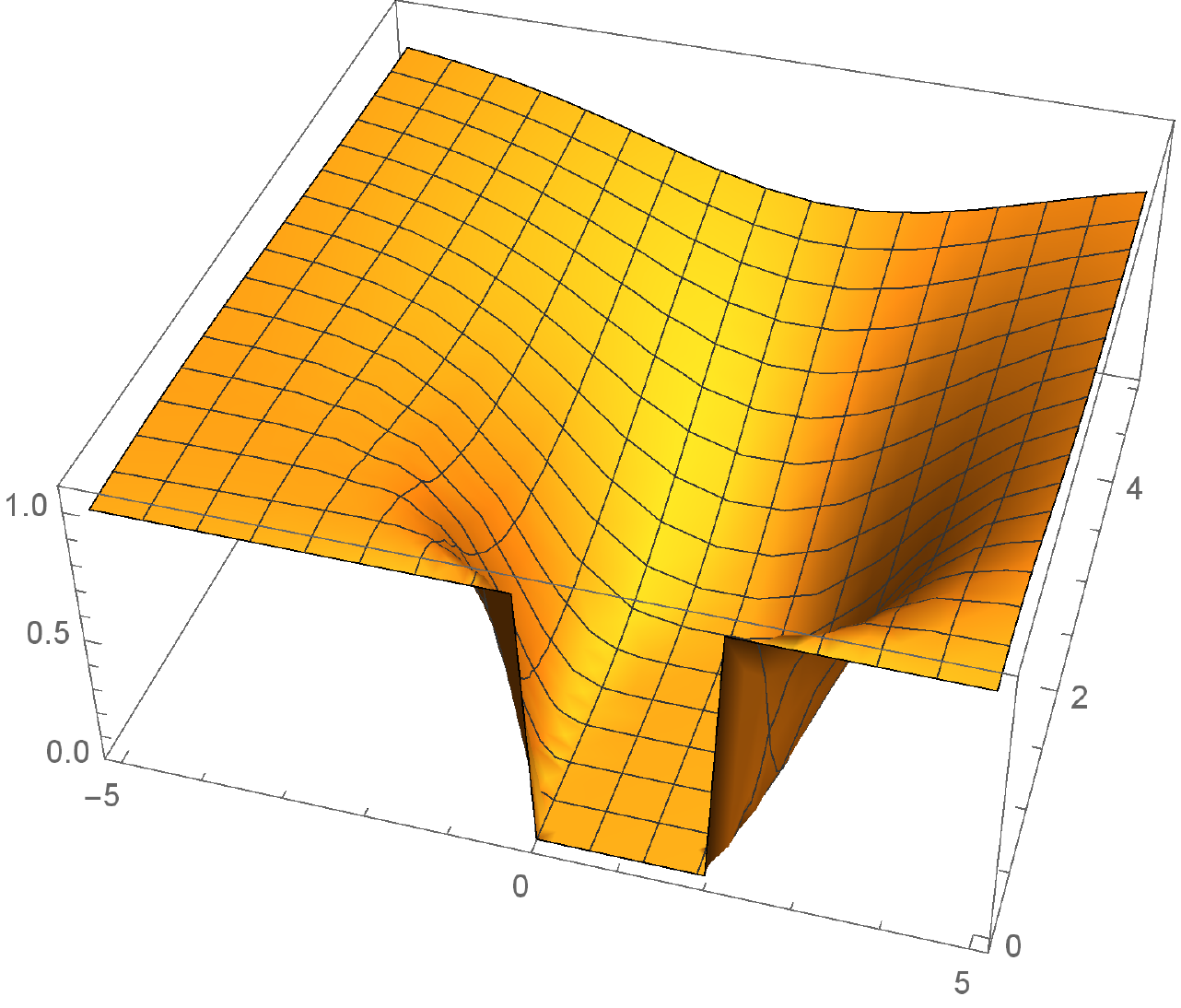}
  \caption{The profiles of $I(\rho,\rho')$ as a function of $x'$ (horizontal axis)
and $\tau'$ (depth axis) for the choice $A=[0,2]$ (i.e.$L=2$).
The left two ones are for a 2d holographic CFT while the right ones for a 2d free scalar CFT.
In the upper two graphs we chose $h=1/2$ and $(x,\tau)=(1,0.1)$ and in the lower two, we chose
$h=10$ and $(x,\tau)=(-1,0.1)$.}
\label{fig:Insidew}
  \end{figure}

In holographic CFTs, we can approximate the correlation functions by regarding the operators are
generalized free fields \cite{ElShowk:2011ag} so that we simply take the Wick contractions of two point functions
(we set $z=z_1$ and $z'=-z_3$):
\be
F(z_1,z_2,z'_3,z'_4)\simeq |z-\bar{z}|^{-4h}\cdot |z'-\bar{z}'|^{-4h}+ |z+\bar{z}'|^{-8h}.  \label{wick}
\ee
The value of $I(\rho,\rho')$ as a function of $w'=x'+i\tau'$ is plotted in the left two graphs in Fig.\ref{fig:Insidew}.
The upper left graph is the case where $w$ is inside the wedge (\ref{outent}) and we have
$I=1$ iff $w=w'$, while $0<I<1$ iff $w\neq w'$, as expected. This shows that we can correctly distinguish the states.
On the other hand, if $w$ is outside the wedge (see the lower left graph), we find $I\simeq 1$ (i.e. indistinguishable)
if $w'$ is also outside, while we have $I\simeq 0$ if $w'$ is inside. We can see that the border is precisely
the CFT counterpart of the entanglement wedge (\ref{outent}). This border gets very sharp when $h\gg 1$ as we are assuming
to justify the geodesic approximated. These behaviors perfectly agree with the distinguishability of bulk states in
the AdS/CFT.

When we calculate the information metric we assume $w\simeq w'$ (or equally $z\simeq z'$).
In this case the first term in (\ref{wick}) dominates when $|z-\bar{z}|\leq |z+\bar{z}|$ and this condition
precisely matches with that for the outside wedge condition. Indeed, if we only keep this first term, we immediately
find $I(\rho,\rho')=1$. On the other hand, when it is inside, the second term is dominant and the result is identical to
the case where $A$ is the total space (i.e. $\rho_A=|\phi(w)\lb\la\phi(w)|$ is pure).

It is instructive to calculate $I(\rho,\rho')$ for non-holographic CFTs. As an example, we consider
a 2d free massless scalar CFT (the scalar field is denoted by $\varphi$) and choose the primary operator to be
$O(w,\bar{w})=e^{i\ap\vp(w,\bar{w})}$, which has the dimension $h=\ap^2/2$. Then we explicitly obtain
\ba
I(\rho,\rho')= \left(\frac{|z+z'|^2|z+\bar{z}||z'+\bar{z}'|}{4|z||z'||z+\bar{z}'|^2}\right)^{4h}.
\ea
The result is plotted as right two graphs in  Fig.\ref{fig:Insidew}.  Clearly in this free CFT, we cannot
find any sharp structure of entanglement wedge as opposed to holographic CFTs, though they have qualitative
similarities (refer to the lower right picture). \\

{\bf 3. Bures Metric}

Now let us calculate the genuine Bures metric when $A$ is a single interval.
We evaluate $\mbox{Tr}[\s{\s{\rho}\rho'\s{\rho}}]$ from
\ba
A_{n,m}=\mbox{Tr}[(\rho^m\rho'\rho^m)^n]. \label{amn}
\ea
via the analytical continuation $n=m=1/2$.
We apply the conformal transformation (we set $k=(2m+1)n$)
\be
z^k=\frac{w}{w-L},  \label{confglk}
\ee
so that the path-integrals for $2mn$ $\rho$s and $n$ $\rho'$s are mapped into that on a single plane.
This leads to
\ba
A_{n,m}\!=\!\frac{\la O^\dagger\!(w_1\!)O(w_2\!)\!\ddd\! O^\dagger\!(w_{2k-1}\!)O  (w_{2k}\!)\lb\!\cdot\! Z^{(k)}}
{\prod_{i=1}^k \la O  ^\dagger(w_{2i-1})O  (w_{2i})\lb\!\cdot\! (Z^{(1)})^k}. \label{ratq}
\ea
Refer to \cite{Nima,Ug} for analogous computations of relative entropy.
After the conformal mapping (\ref{confglk}), we find
\ba
&& A_{n,m}=\prod_{i=1}^{2k}\left|k^{-1}(z_i)^{1-k}\right|^{2h}\cdot \prod_{j=1}^k |(z_{2j-1})^k-(z_{2j})^k|^{4h}  \no
&& \times \la O  ^\dagger(z_1)O  (z_2)\ddd O  ^\dagger(z_{2k-1})O(z_{2k})\lb\cdot
\f{Z^{(k)}}{(Z^{(1)})^k}.  \label{cosing}
\ea
Note that we have
\ba
&& z_1=\left(\frac{-x-i\tau}{L-x-i\tau}\right)^{1/k},\ \ \ z_2(=\bar{z}_1)=\left(\frac{-x+i\tau}{L-x+i\tau}\right)^{1/k}, \no
&& z_{2s+1}=e^{\frac{2\pi i}{k}s}z_1,\ \ \ z_{2s+2}=e^{\frac{2\pi i}{k}s}z_2,\ \ \ (s=1,2,\ddd,k-1).\nonumber
\ea

Let us evaluate $A_{n,m}$ in holographic CFTs, using the generalized free field approximation.
We take $w\simeq w'$ to calculate the Bures metric.  When $w$ and $w'$ are outside of the entanglement wedge (\ref{outent}), or 
equally $|z_{2j-1}-z_{2j}|<|z_{2j-2}-z_{2j-1}|$, the $2k$ point function is approximated as
\ba
&& \la O  ^\dagger(z_1)O  (z_2)\ddd O  ^\dagger(z_{2k-1})O  (z_{2k})\lb\no
&& \simeq  \prod_{j=1}^k \la O  ^\dagger(z_{2j-1})O  (z_{2j})\lb \simeq  \prod_{j=1}^k |z_{2j-1}-z_{2j}|^{-4h},
\ea
and this leads to the trivial result $A_{n,m}=1$, leading the vanishing Bures metric
$dD_B^2= 0$. This agrees with the AdS/CFT expectation that $\rho_A$ cannot distinguish two different bulk
excitations outside of entanglement wedge.

On the other hand, when $w$ and $w'$ are inside of the entanglement wedge (\ref{outent}), or 
equally $|z_{2j-1}-z_{2j}|>|z_{2j-2}-z_{2j-1}|$, we can approximate as
\ba
&& \la O^\dagger(z_1)O(z_2)\ddd O^\dagger(z_{2k-1})O(z_{2k})\lb\no
&& \simeq  \prod_{j=1}^k \la O^\dagger(z_{2j-2})O(z_{2j-1})\lb \simeq  \prod_{j=1}^k |z_{2j-2}-z_{2j-1}|^{-4h}.\nonumber
\ea
In the limit $n\to 1/2$ and $m\to 1/2$, this leads to
\ba
&& A_{1/2,1/2}= |w-\bar{w}|^{2h}|w'-\bar{w}'|^{2h}|w'-\bar{w}|^{-4h}, \no  \label{www} \no
&& dD_B^2= \frac{h}{\tau^2}(dx^2+d\tau^2).  \label{wwww}
\ea
This Bures metric for $\rho_A(w,\bar{w})$ coincides with that for the pure state (\ref{hypp}) and reproduces
the bulk AdS metric on the time slice $\tau=0$.

Similarly, in a 2d holographic CFT with a circle compactification $x\sim x+2\pi$, we obtain the Bures metric
\ba
dD_B^2=\frac{h}{\sinh^2\tau}(d\tau^2+dx^2),  \label{bgsl}
\ea
if $w$ is inside the wedge.  In a 2d holographic CFT at finite temperature $T$, the Bures metric is computed as
\ba
dD_B^2=h\frac{(2\pi T)^2}{\sin^2\left(2\pi T\tau\right)}(d\tau^2+dx^2).  \label{btzbrm}
\ea
if $w$ is inside the wedge. These metrics agree with those on the time slice $\tau=0$ of global AdS$_3$ and BTZ black hole,
by projecting the point $(x,\tau)$ at the AdS boundary into the time slice along each geodesic.
In summary, our CFT calculations for these  setups show that in holographic CFTs, we can distinguish two excitations if they are inside the entanglement wedge.

It is intstructive to calculate the Bures metric in the 2d massless free scalar CFT for the primary
$O=e^{i\ap\vp}$. For $\ap=1$, we find the following analytical result:
\ba
&& A_{\f12,\f12}\!=\!\frac{(\s{z}+\s{z'})(\s{\bar{z}}+\s{\bar{z'}})(\s{z}+\s{\bar{z}})(\s{z'}+\s{\bar{z'}})}{4\s{|z||z'|}(\s{z}+\s{\bar{z}'})(\s{\bar{z}}+\s{z'})},\no
&& dD_B^2=-\frac{L^2(dw)^2}{16w^2(L-w)^2}-\frac{L^2(d\bar{w})^2}{16\bar{w}^2(L-\bar{w})^2}\no
&&\ \ \ \ \ +\frac{L^2(dw)(d\bar{w})}{2|w||w-L|\left(\s{\bar{w}(w-L)}+\s{w(\bar{w}-L)}\right)^2}.
\ea
Note that we cannot find any sharp structure of entanglement wedge as opposed to
the holographic CFT. However, in the limit $\tau\to 0$, we find the metric $ds^2\simeq \frac{h}{\tau^2}(d\tau^2+dx^2)$
for $0\leq x\leq L$.\\

{\bf 4. Double Interval Case}

Finally, we take the subsystem $A$ to be a union of two disconnected intervals $A_1$ and $A_2$, which are  parameterized as
$A_1=[0,s]$ and $A_2=[l+s,l+2s]$, without losing generality. We conformally map the $w$-plane with two slits along $A_1$ and $A_2$
into a $z-$cylinder via (see e.g. \cite{Raj})
\ba
z=f(w)=\frac{J(\kappa^2)}{2}-\frac{J(\kappa^2)}{2K(\kappa^2)}\mbox{sn}^{-1}(\ti{w},\kappa^2),\ \label{zfw}
\ea
where we introduced
\ba
&& \ti{w}=\frac{2}{l}\left(w-s-\frac{l}{2}\right),\ \ \ J(\kappa^2)=2\pi\frac{K(\kappa^2)}{K(1-\kappa^2)},\no
&& K(\kappa^2)=\int^1_0\frac{dx}{\s{(1-x^2)(1-\kappa^2 x^2)}},\ \ \ \kappa=\frac{l}{l+2s}.\nonumber
\ea
The function $\mbox{sn}^{-1}(\ti{w},\kappa^2)$ is the Jacobi elliptic function:
\be
\mbox{sn}^{-1}(\ti{w},\kappa^2)=\int^{\ti{w}}_0 \frac{dx}{\s{(1-x^2)(1-\kappa^2 x^2)}}.
\ee
It is useful to note the relation $\mbox{sn}^{-1} (\ti{w},0)=\arcsin(\ti{w})$.

\begin{figure}
  \centering
  \includegraphics[width=5cm]{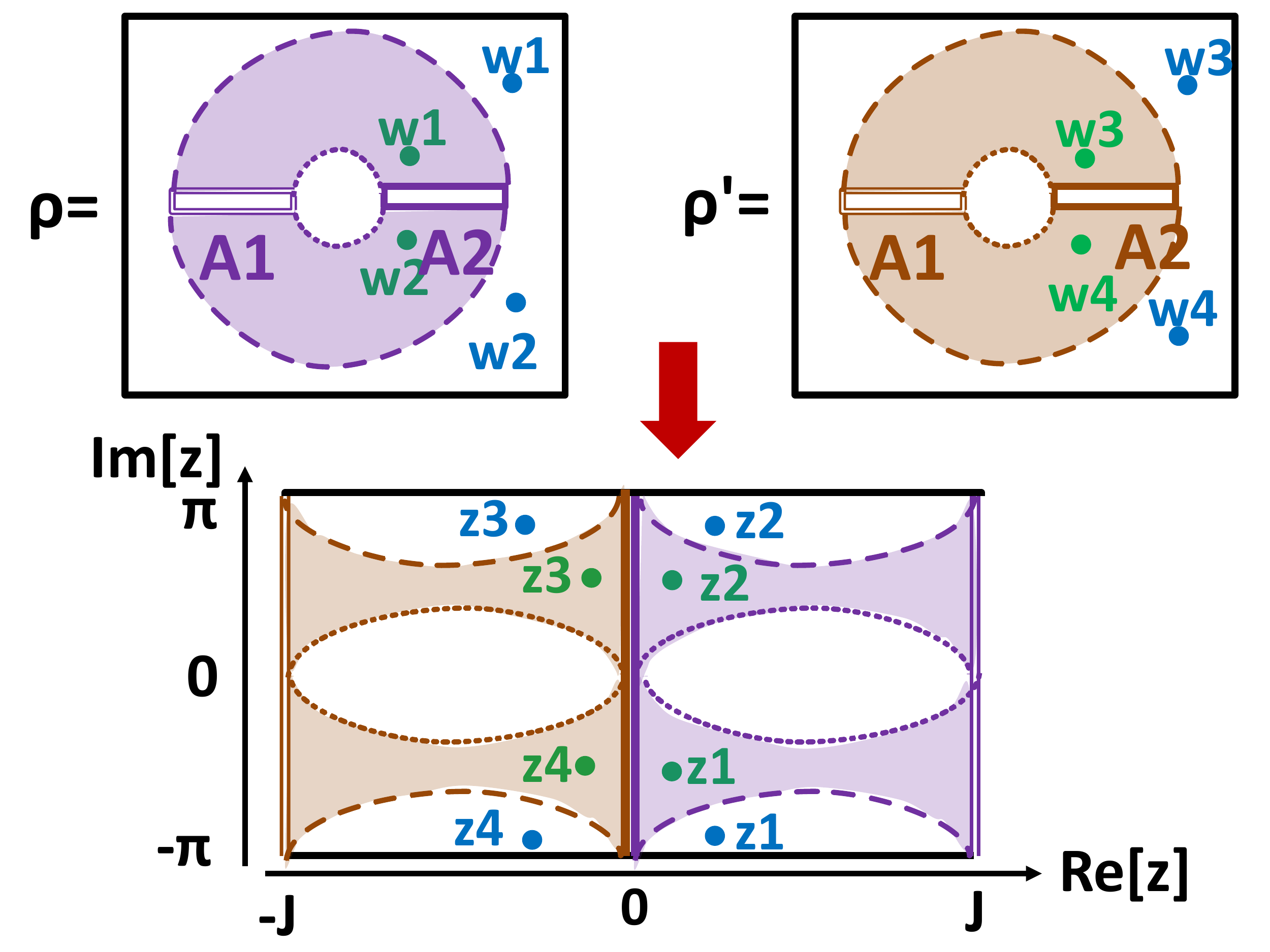}
  \caption{A sketch of  conformal transformation for Tr$[\rho_A\rho'_A]$
in the double interval case. We assumed the phase (i), where
the entanglement wedge is connected, as depicted by the colored region.
The lower picture describes the geometry after the transformation and is given by a torus by identifying the edges.
Blue (or Green) points are outside (or inside) of  $M_A$. }
\label{fig:double}
  \end{figure}

We can calculate $I(\rho,\rho')$ using the map (\ref{zfw}) both for $\rho_A(w,\bar{w})$ and $\rho_A(w',\bar{w}')$
and the formula (\ref{formi}). The two $w-$planes are mapped into a torus, described by the
$z-$plane with the identification Re$[z]\sim$Re$[z]+2J$ and Im$[z]\sim$Im$[z]+2\pi$, as depicted in Fig.\ref{fig:double}.

In holographic CFTs, we need to distinguish two phases depending on the moduli of the torus:
\ba
&& (i)\  \mbox{Connected phase}: J<\pi \ \ \mbox{or equally}\ \kappa < 3-2\s{2} ,\no
&& (ii)\ \mbox{Disconnected phase}: J>\pi \ \ \mbox{or equally}\ \kappa > 3-2\s{2} .\nonumber
\ea
We can confirm the phase $(i)$ (or  $(ii)$) coincides with the case in the gravity dual where the entanglement wedge gets connected
(or disconnected), and the circle Re$[z]$ (or  Im$[z]$) shrinks to zero size in the bulk, respectively.
This is the standard Hawking-Page transition \cite{HP} and agrees with the large $c$ CFT analysis \cite{He}.
The holographic two point functions on the torus in each phase behaves like
\ba
&& \la O^\dagger\!  (z,\!\bar{z})O\!(z',\!\bar{z}')\lb_{(i)}\!\simeq\! \mbox{Max}_{n_1\in Z}
\left|\sin\left(\frac{\pi(z\!+\!2\pi in_1\!-\!z')}{2J}\right)\right|^{-4h}\!\!,\no
&& \la O^\dagger\!(z,\!\bar{z})O\!(z',\!\bar{z}')\lb_{(ii)}\!\simeq\! \mbox{Max}_{n_2\in Z}
\left|\sinh\left(\frac{(z\!+\!2J n_2\!-\!z')}{2}
\right)\right|^{-4h}\!\!.\nonumber
\ea

Let us estimate  $4-$point functions $F$ in (\ref{formi}) by the generalized free field
prescription, where we again assume $w\simeq w'$.  There are two contributions:
the trivial Wick contraction and the non-trivial one as in (\ref{wick}).
The trivial one leads to $I(\rho,\rho')=1$, which tells us that
we cannot distinguish the two nearby states. Therefore we again find that the entanglement wedge
corresponds to the region where non-trivial contractions get dominant.

\begin{figure}
  \centering
\includegraphics[width=2.2cm]{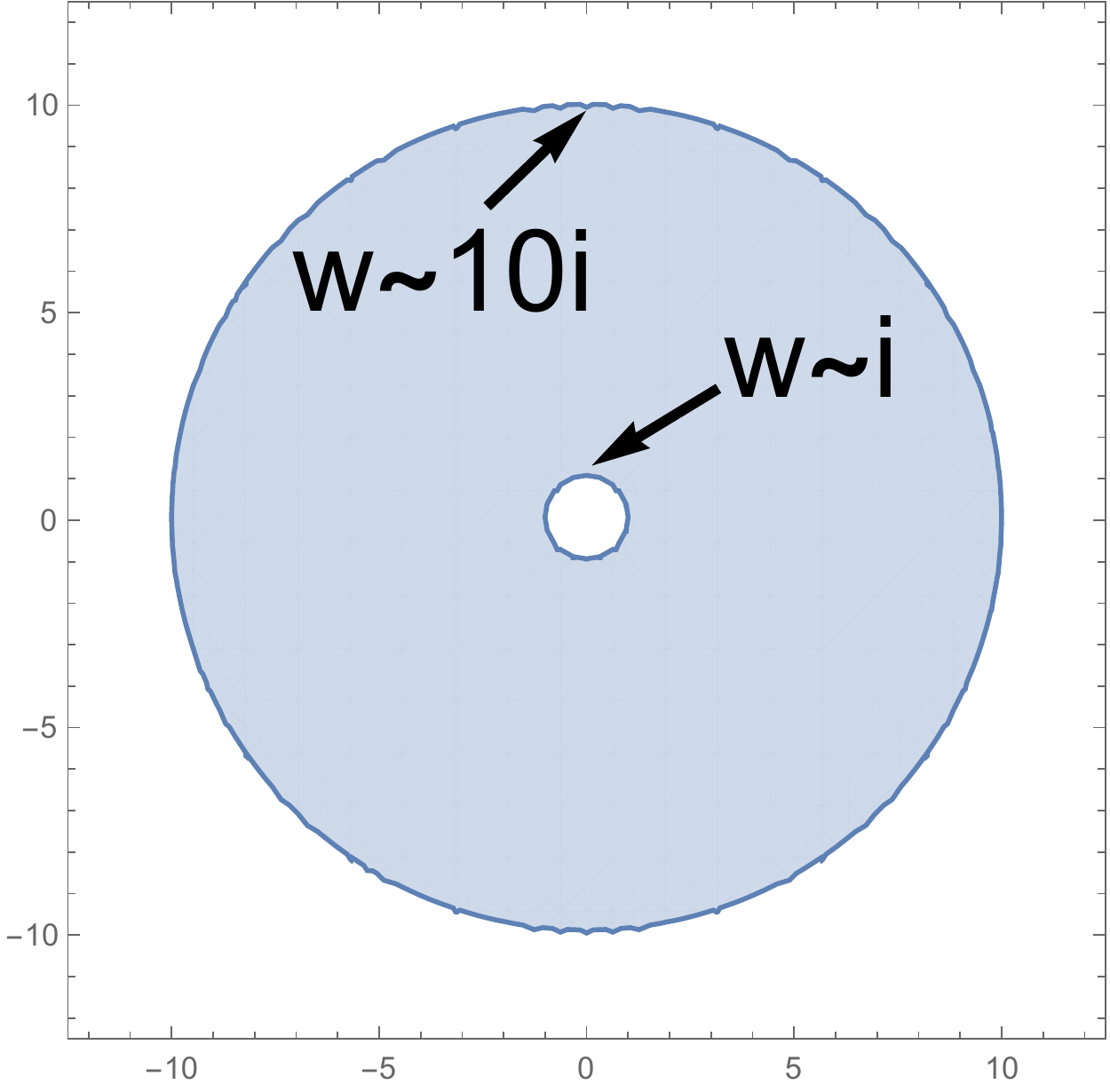}
  \includegraphics[width=2.2cm]{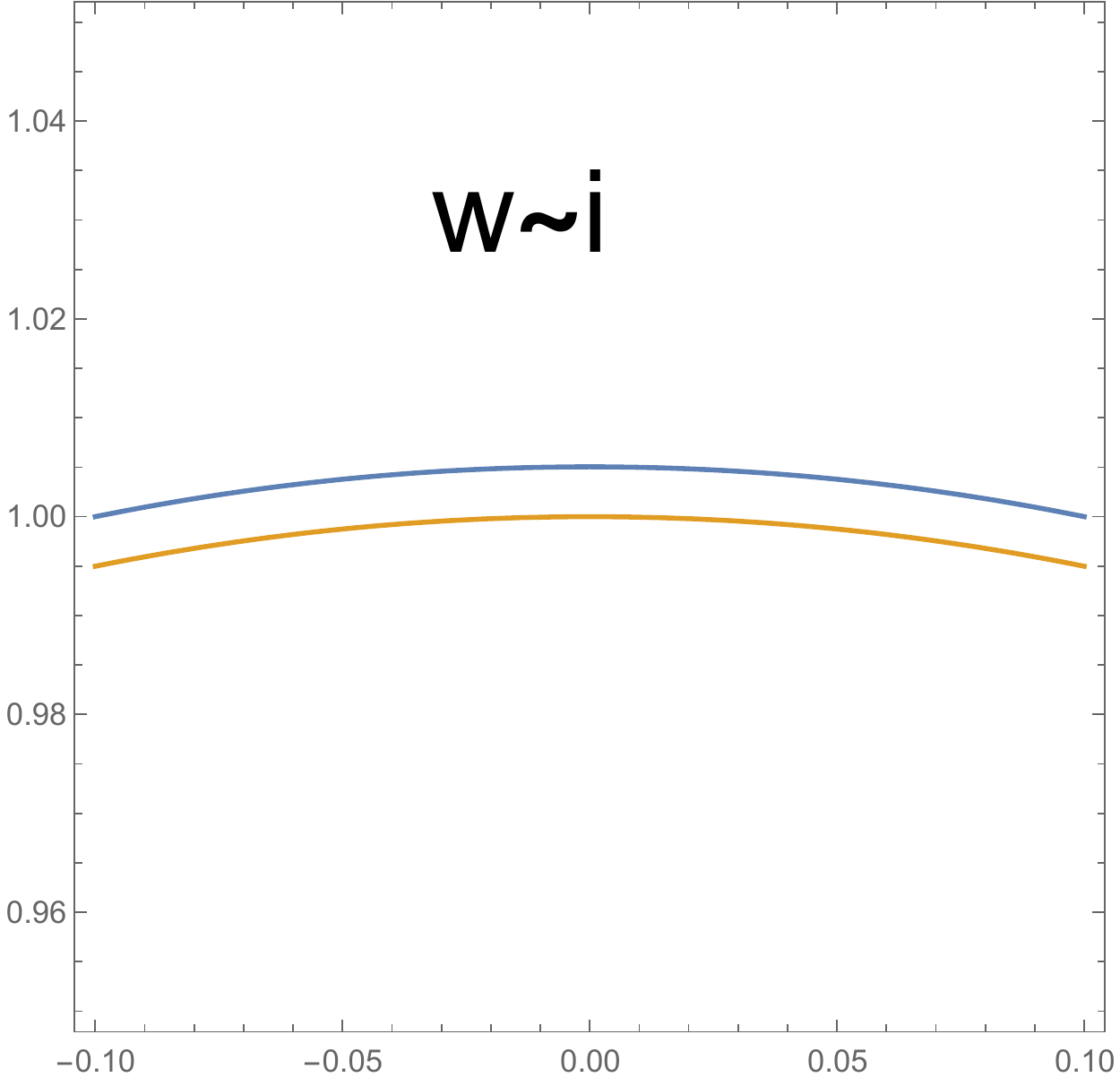}
\includegraphics[width=2.2cm]{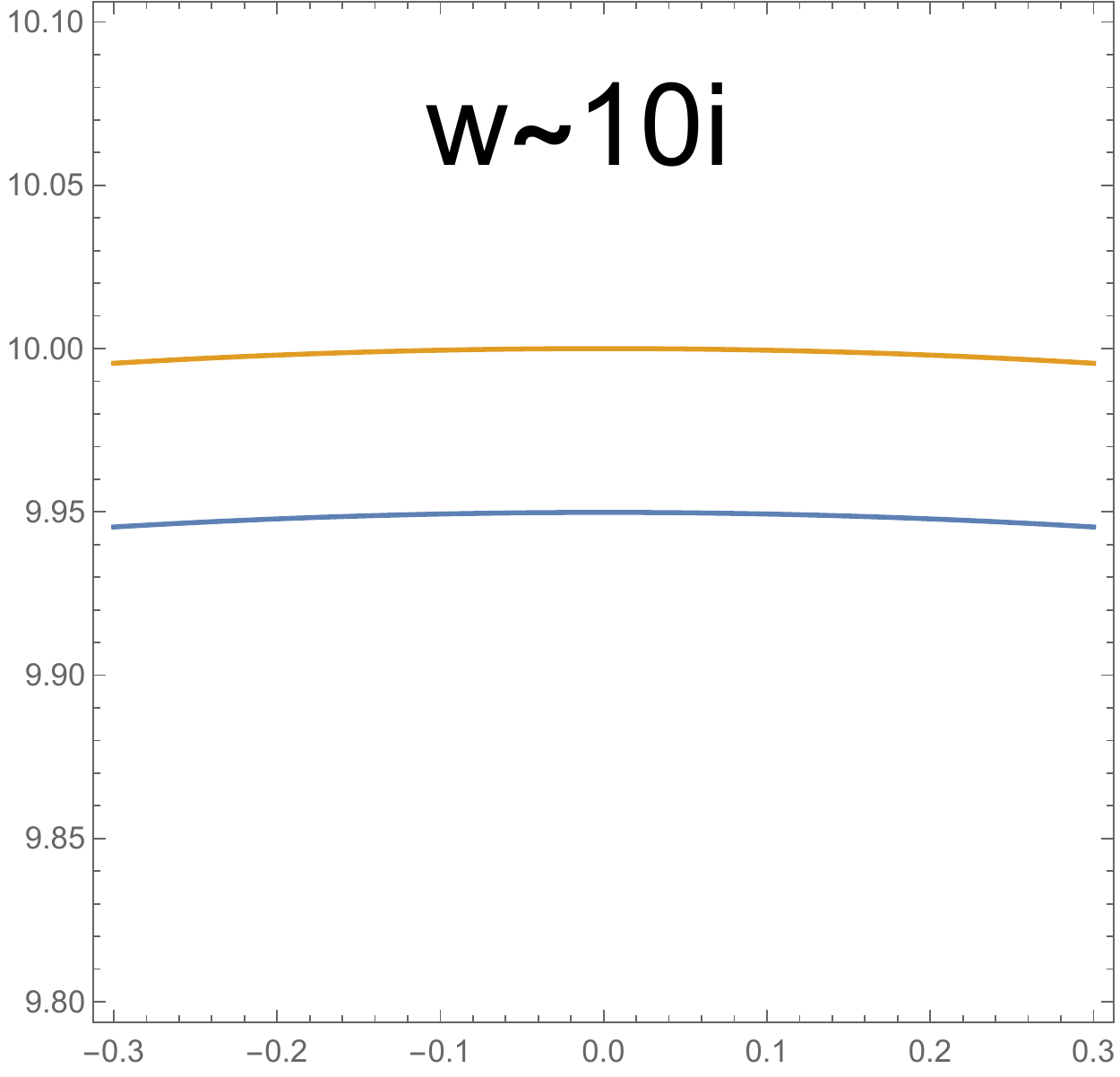}
  \includegraphics[width=2.2cm]{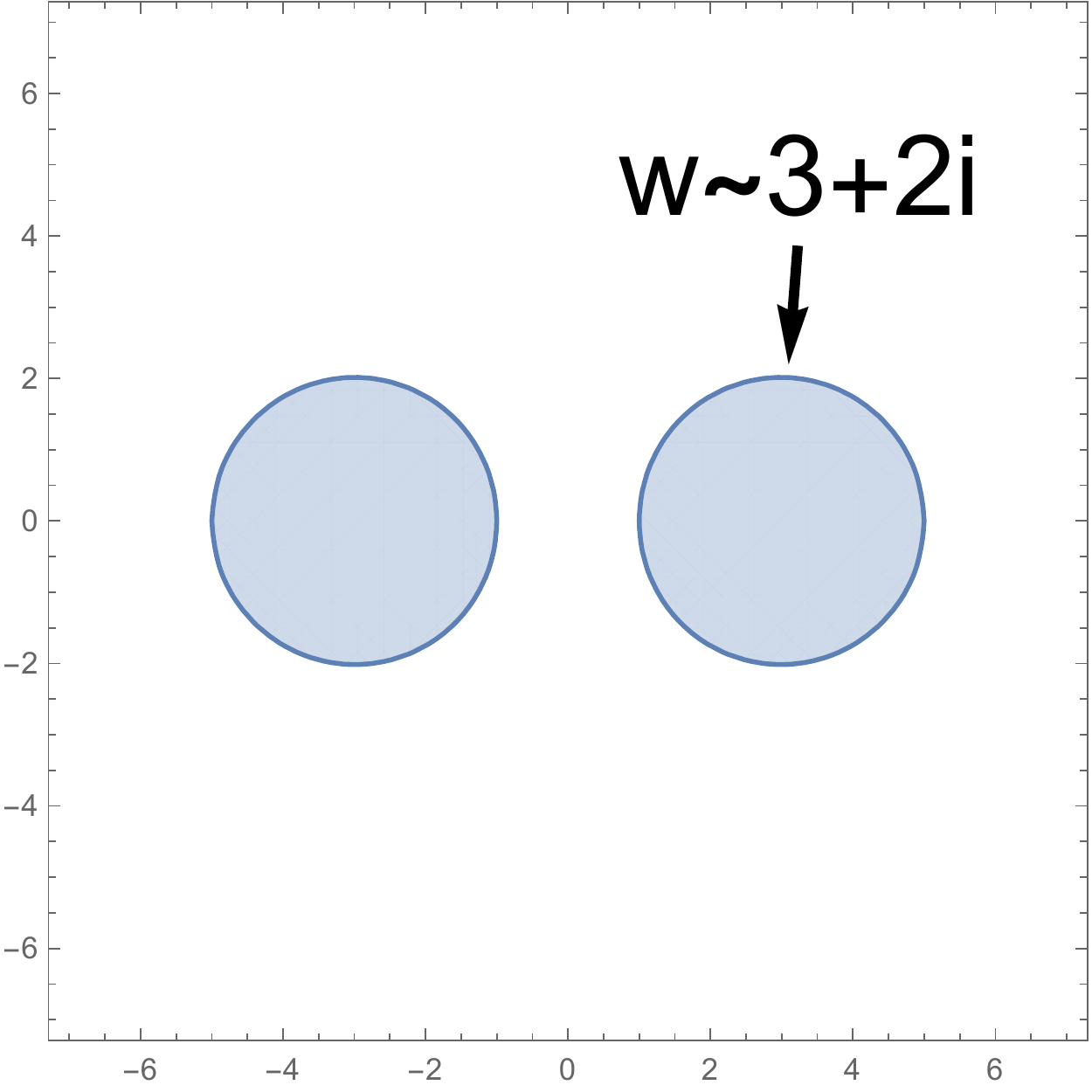}
 \includegraphics[width=2.2cm]{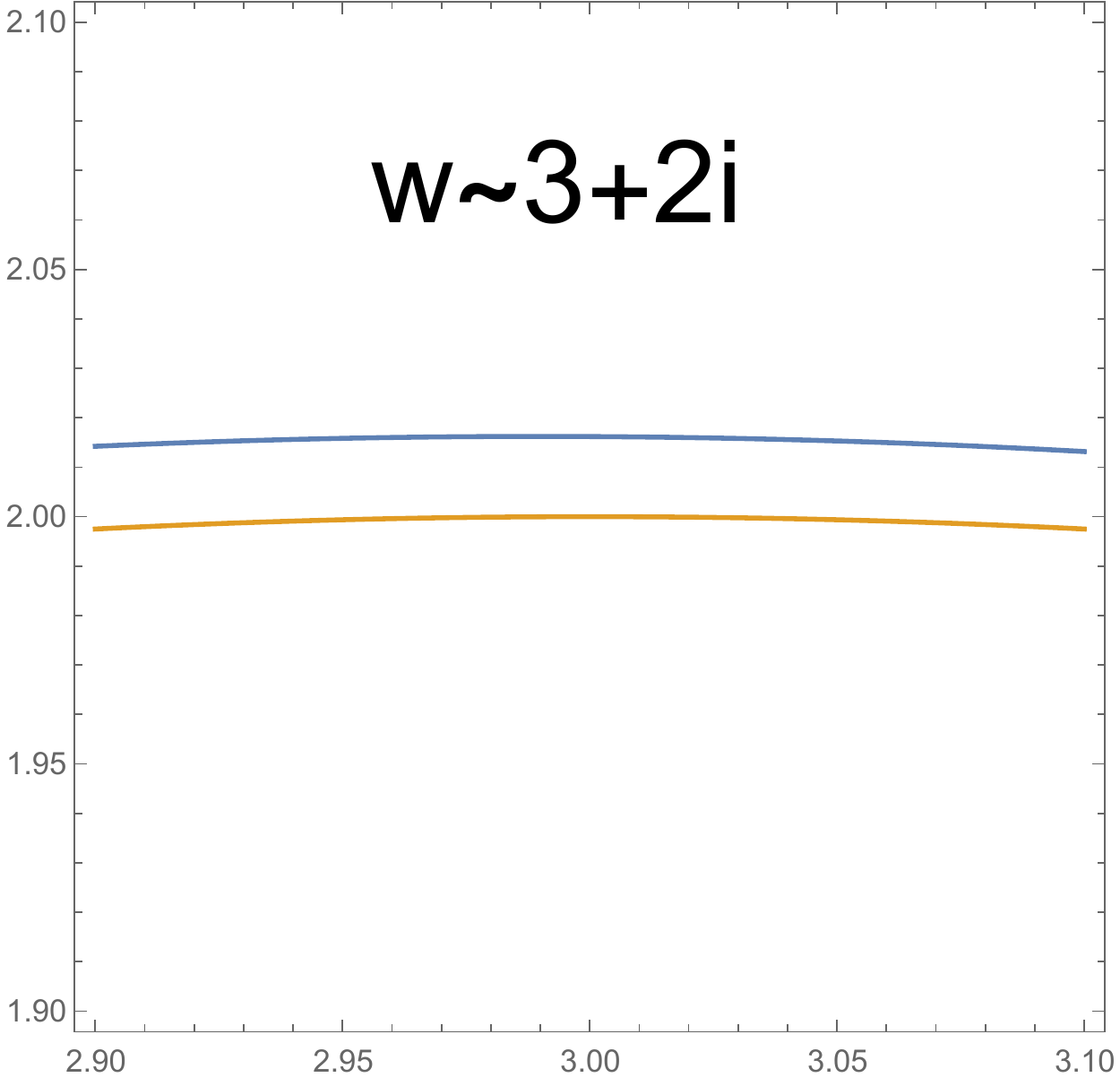}
  \caption{The plots of the locations of the operator insertion on $\ti{w}-$plane where the non-trivial Wick contraction
is favored (blue colored regions in the left pictures).
The upper pictures are for  $\kappa=0.1$ where $M_A$ is connected, while the lower ones are for $\kappa=0.2$ where $M_A$ is disconnected.
In the upper middle and right picture, blue curves are the borders between the non-trivial and trivial contraction, while
orange curves describe the borders of the entanglement wedge. The same is true for the lower right picture.}
\label{fig:discondd}
  \end{figure}

The non-trivial Wick contraction is dominant when
\ba
\mbox{Min}_{n_1\in Z}\left|\sin\left(\frac{\pi}{2J}(z_2\!-\!z_1\!-\!2n_1\pi i)\right)\right|
\geq \left|\sin\left(\frac{\pi}{2J}(z_3\!-\!z_2)\right)\right|,\nonumber
\ea
in the connected case, and when
 \ba
\left|\sinh\left(\frac{z_2-z_1}{2}\right)\right| \geq
\mbox{Min}_{n_2\in Z}\left|\sinh\left(\frac{z_2-z_3-2n_2J}{2}\right)\right|, \nonumber
\ea
in the disconnected case.
We plotted these regions in terms of the coordinate $\ti{w}$ of (\ref{zfw}) in Fig. \ref{fig:discondd}.

\begin{figure}
  \centering
\includegraphics[width=3cm]{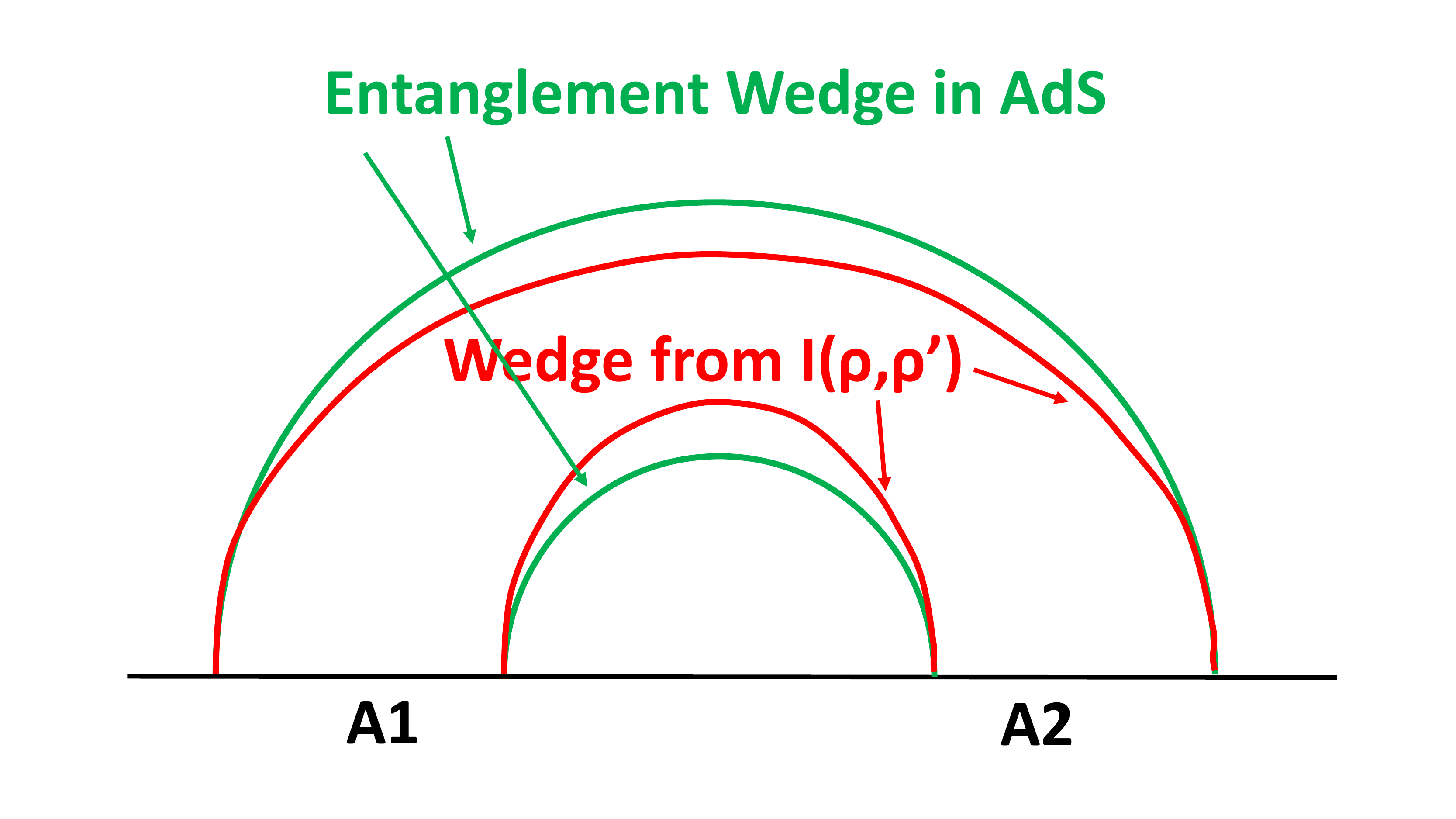}
  \includegraphics[width=3cm]{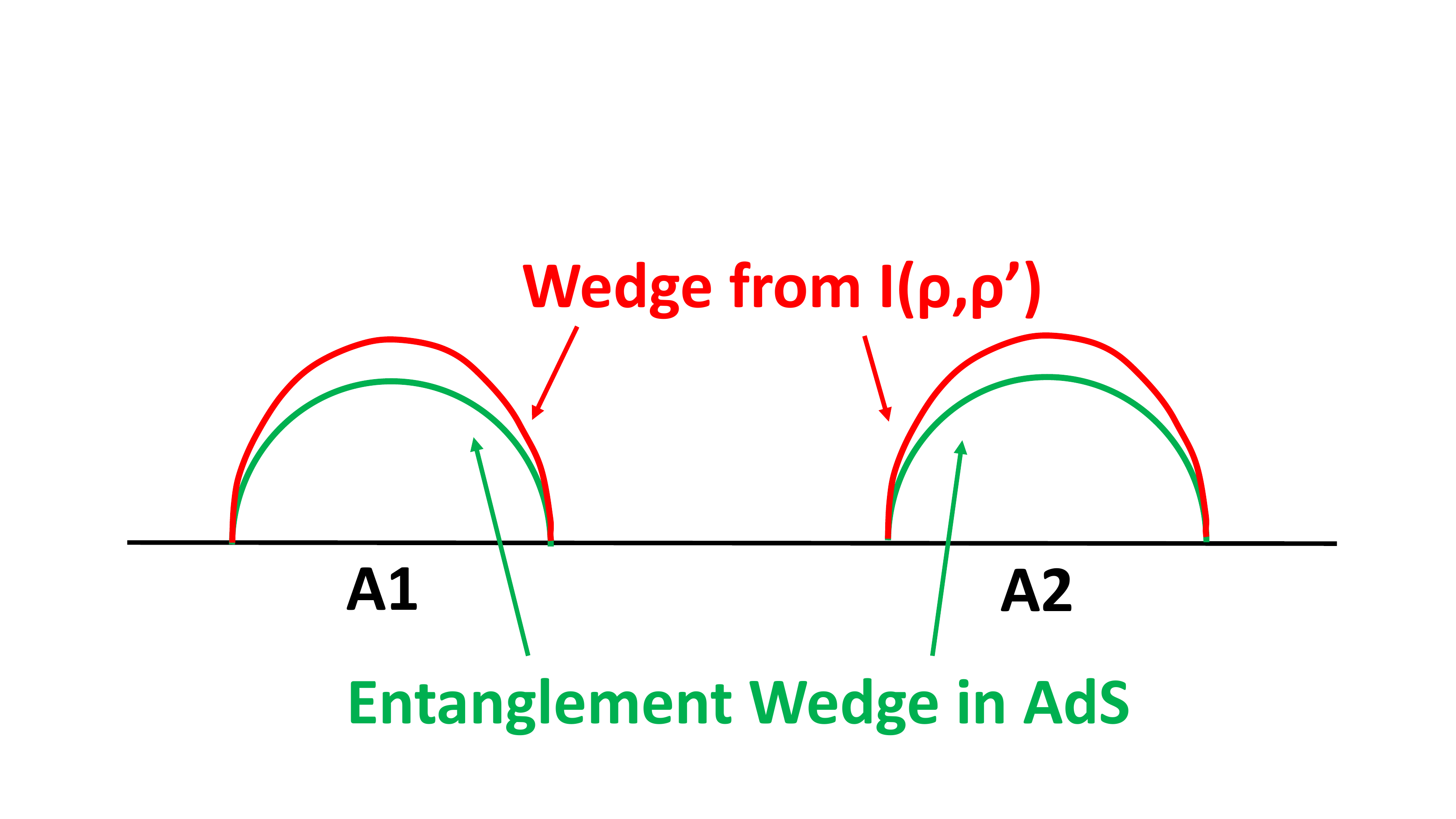}
  \caption{Sketches which exaggerate small deviations between the wedges from $I(\rho,\rho')$ and the actual entanglement wedges.}
\label{fig:deviation}
  \end{figure}

In both cases, the regions are very close to the true entanglement wedges, respecting
the expected connected or disconnected geometry (note our holographic relation in Fig.\ref{fig:EW}). The deviation is always
within a few percent, depicted in Fig.\ref{fig:discondd}. This small deviation arises as
the correct distinguishability should be measured by the Bures metric.
Our analysis using $I(\rho,\rho')$ only gives an approximation, much like the Renyi
entropy compared with the von-Neumann entropy. As sketched in Fig.\ref{fig:deviation}, 
this wedge from $I(\rho,\rho')$ obeys the following rules: (a) the wedge for $A_1\cup A_2$ is larger than the union of the wedges 
for $A_1$ and $A_2$ and (b) the wedge for $A$ is the complement to the one for $A^c$.

Thus it would be ideal  if we can calculate the genuine Bures distance $D_B(\rho,\rho')$ in the double interval case.
This is very complicated as the trace $\mbox{Tr}[(\rho^m\rho'\rho^m)^n]$ corresponds to a
partition function on a genus $n(2m+1)-1$  Riemann surface. However, since we will finally take $n=m\to 1/2$ limit (genus 0 limit),
it might not be surprising to obtain the expected metric (\ref{hypp})
which coincides with the case where $A$ is the total space.\\

{\bf 5. Discussions}

In this article, we presented a general mechanism how entanglement wedges emerge from holographic CFTs and gave
several successful examples. One important furture problem is to repeat the same procedure by using the genuine localized
operator in the bulk \cite{HKLL} or the state \cite{MNSTW}, which may give us more refined results.
Another interesting
direction will be to extend this construction to the higher dimensional AdS/CFT.  Moreover, it may be useful to consider
other distance measures such as trace distances \cite{ZRC}. It is also intriguing to explore the relationship between our approach and the path-integral optimization \cite{Caputa:2017urj}.
It might also be fruitful to consider connections between our results and the recent proposals for entanglement wedge cross sections \cite{UT,Nguyen:2017yqw,Kudler-Flam:2018qjo,CMTU,Tamaoka:2018ned,Dutta:2019gen}. We would like to come back to
 these problems soon later \cite{STU}.\\

{\bf Acknowledgements} We thank Pawel Caputa, Veronika Hubeny, Henry Maxfield, Mukund Rangamani, Hiroyasu Tajima and Kotaro Tamaoka for useful conversations. TT is supported by the Simons Foundation through the ``It from Qubit'' collaboration. TT is supported by JSPS Grant-in-Aid for Scientific Research (A) No.16H02182 and by JSPS Grant-in-Aid for Challenging Research (Exploratory) 18K18766. TT is also supported by World Premier International Research Center Initiative (WPI Initiative) from the Japan Ministry of Education, Culture, Sports, Science and Technology (MEXT). KU is supported by Grant-in-Aid for JSPS Fellows No.18J22888. We are grateful to the long term workshop ”Quantum Information and String Theory” (YITP-T-19-03) held at Yukawa Institute for Theoretical Physics, Kyoto University and participants for useful discussions. TT thanks very much the workshop ``Quantum Information in Quantum Gravity V,'' held in UC Davis, where this work was presented.


\end{document}